\documentclass[reprint,prl,superscriptaddress,showpacs,twocolumn]{revtex4}
\usepackage{units}
\usepackage{amsmath}
\usepackage{amssymb}
\usepackage{graphicx}
\usepackage{bm}
\usepackage{microtype,color}

\newcommand{\be}{\begin{equation}}
\newcommand{\ee}{\end{equation}}
\newcommand{\bea}{\begin{eqnarray}}
\newcommand{\eea}{\end{eqnarray}}

\begin{document}

\title{Rotating optical microcavities with broken chiral symmetry}

\author{Raktim Sarma}
\affiliation{Department of Applied Physics, Yale University, New Haven, CT, 06520, USA}
\author{Li Ge}
\affiliation{\textls[-20]{Department of Engineering Science and Physics, College of Staten Island, CUNY, Staten Island, NY 10314, USA}}
\affiliation{\textls[-20]{The Graduate Center, CUNY, New York, NY 10016, USA}}
\author{Jan Wiersig}
\affiliation{Institut f{\"u}r Theoretische Physik, Universit{\"a}t Magdeburg, Magdeburg, Postfach 4120, Germany}
\author{Hui Cao}
\email{hui.cao@yale.edu}
\affiliation{Department of Applied Physics, Yale University, New Haven, CT, 06520, USA}
%$\footnote{e-mail:hui.cao@yale.edu}$

\date{\today}

\begin{abstract}
We demonstrate in open microcavities with broken chiral symmetry, quasi-degenerate pairs of co-propagating modes in a non-rotating cavity evolve to counter-propagating modes with rotation. The emission patterns change dramatically by rotation, due to distinct output directions of CW and CCW waves. By tuning the degree of spatial chirality, we maximize the sensitivity of microcavity emission to rotation. The rotation-induced change of emission is orders of magnitude larger than the Sagnac effect, pointing to a promising direction for ultrasmall optical gyroscopes.

\end{abstract}

\maketitle

Light propagation in rotating systems has been studied as one of the most fundamental problems of electromagnetics \cite{Shiyozawa1,EJPost1,Harayama1,Steinberg1,Steinberg2,Joannopoulos1}.
Rotation-induced phase difference or frequency splitting between counter-propagating beams in a loop, the so-called Sagnac effect, has been widely used for rotation sensing \cite{EJPost1,Scully,Ciminell1,Sorrentino1,Scheur1}.
Recently microcavity lasers have been explored for ultrasmall optical gyroscopes \cite{Harayama1,Scheur2,Sarma1}.
Since the frequency splitting is proportional to the cavity radius and becomes very small in microcavities, rotation-induced changes in other properties have been investigated \cite{Scheur2,Sarma1}.
For example, the cavity quality factor, which determines lasing threshold and output power, is shown to be more sensitive to rotation, pointing to possible new direction for rotation sensing.
However, it is not yet known what cavity geometry, which type of resonances, and what optical property are most sensitive to rotation.

In addition to potential applications, recent studies on rotating optical microcavities have deepened the fundamental understanding of light propagation in the rotating frame \cite{Harayama1,Harayama2,Harayama3}.
In a two-dimensional (2D) cavity of shape deformed from a circle, the clockwise (CW) and counter-clockwise (CCW) propagating waves are usually coupled to form standing-wave resonances with non-degenerate frequencies. The absence of CW and CCW traveling-wave resonances at degenerate frequencies results in a threshold for Sagnac effect, i.e. the frequency shift due to rotation occurs only when the rotation is fast enough \cite{Harayama1,Harayama2}.
All the cavities that have been studied so far possess the chiral symmetry, namely, $r(-\theta) = r(\theta)$, where $r(\theta)$ describes the cavity boundary in the polar coordinates.
In the absence of rotation, the CW and CCW waves have equal contributions to each resonance.
However, slight shape deformation, introduced unintentionally during the fabrication of a circular cavity, may break the chiral symmetry. In recent years, microcavities with (intentionally) broken chiral symmetry have been investigated, e.g., the spiral-shaped disk \cite{Spiral1,Spiral2,Spiral3,Spiral4, Spiral5}, the asymmetric lima\c{c}on cavity \cite{Wiersig1}.
Such cavities, termed ``chiral cavities'' here, support pairs of co-propagating modes with a preferred sense of rotation even in the stationary frame~\cite{Spiral5}.
It is not clear whether the lack of counter-propagating waves in the cavity resonances would affect the Sagnac effect, which relies on the coexistence of traveling waves in both directions.

In this Letter, we investigate 2D chiral microcavities in the rotating frame and show that a quasi-degenerate pair of co-propagating modes evolve to counter-propagating ones at sufficient high rotation speed. The frequency shift due to rotation is similar to that in a symmetric cavity. Surprisingly, the emission pattern of a chiral microcavity is changed dramatically by rotation, owing to distinct output directions for CW and CCW waves. We are able to tune the degree of chirality of the cavity shape without spoiling the quality factor \cite{SI}. The maximal chirality leads to the largest difference in CW and CCW output directions, making the emission pattern most sensitive to rotation. The numerical simulation confirms that the rotation-induced change in emission pattern can be orders of magnitude higher than the Sagnac frequency splitting. These results lead to potential application of chiral microcavities to ultrasmall optical gyroscopes.

In a non-rotating chiral cavity, the coupling between CW and CCW propagating waves is asymmetric, leading to the formation of resonances with spatial chirality \cite{Wiersig1,Wiersig2}.
A general description is given by a non-Hermitian $2 \times 2$ effective Hamiltonian~\cite{Wiersig1}

\begin{equation}
H_0 = \left(\begin{array}{cc}
\omega_0 & 0 \\
0 & \omega_0\\
\end{array}\right)
+
\left(\begin{array}{cc}
\Gamma & V \\
\eta V^* & \Gamma\\
\end{array}\right)
\end{equation}

where $\omega_0$ is the frequency of the unperturbed CCW and CW wave components. Their coupling leads to an overall frequency shift $\Gamma$, and asymmetric transition elements $V$ and $\eta V^*$, where $|\eta| < 1$ represents the degree of asymmetry.

Next we introduce rotation to the Hamiltonian. In a circular cavity, the rotation does not couple CW and CCW traveling waves, so the off-diagonal terms of $H$ remain unchanged, but the diagonal terms are modified as the frequency of the CCW (CW) traveling wave component is changed by $\Delta$ ($-\Delta$), where $\Delta$ is linearly proportional to the rotation frequency $\Omega$~\cite{Woerdman}. Assuming this also holds for a deformed cavity, the resulting Hamiltonian is
\begin{equation}
H = H_0 + \left(\begin{array}{cc}
\Delta & 0 \\
0      & -\Delta\\
\end{array}\right)
= \left(\begin{array}{cc}
\omega_0+ \Gamma +\Delta & V \\
\eta V^* & \omega_0 + \Gamma - \Delta\\
\end{array}\right) \ .
\end{equation}

We diagonalize the Hamiltonian to obtain eigenfrequencies and eigenvectors \cite{SI}.
The frequency splitting is $\Delta\omega = 2\sqrt{\eta |V|^2 + \Delta^2}$.
For simplicity, we set $\Delta = \Omega$.
In a symmetric cavity ($\eta = 1$), with small rotation the eigenmodes remain standing-wave modes with equal weights of CCW and CW components, and their frequency difference is barely changed by rotation (Fig. 1).
When the rotation speed is sufficiently high, one mode evolves to a CCW traveling-wave mode, the other one to a CW traveling-wave mode; and their frequency difference starts to grow significantly with $\Omega$.
Hence, the frequency splitting at $\Omega = 0$, as a result of CW and CCW wave coupling in a deformed cavity, causes a dead zone for the Sagnac effect, as predicted in Ref.\cite{Harayama1}.
In a chiral cavity ($\eta = 0.1$), the evolution of frequency splitting with rotation is identical to the symmetric cavity since $\eta |V^2|$ is kept the same [Fig. 1(a)].
Although without rotation both modes are composed mainly of CCW traveling waves, one of them transforms  into a CW traveling wave mode by rotation.
Therefore, in terms of the Sagnac effect, the chiral cavity behaves similar to the symmetric cavity.

\begin{figure}[htbp]
\centering
\includegraphics[width=1\linewidth]
{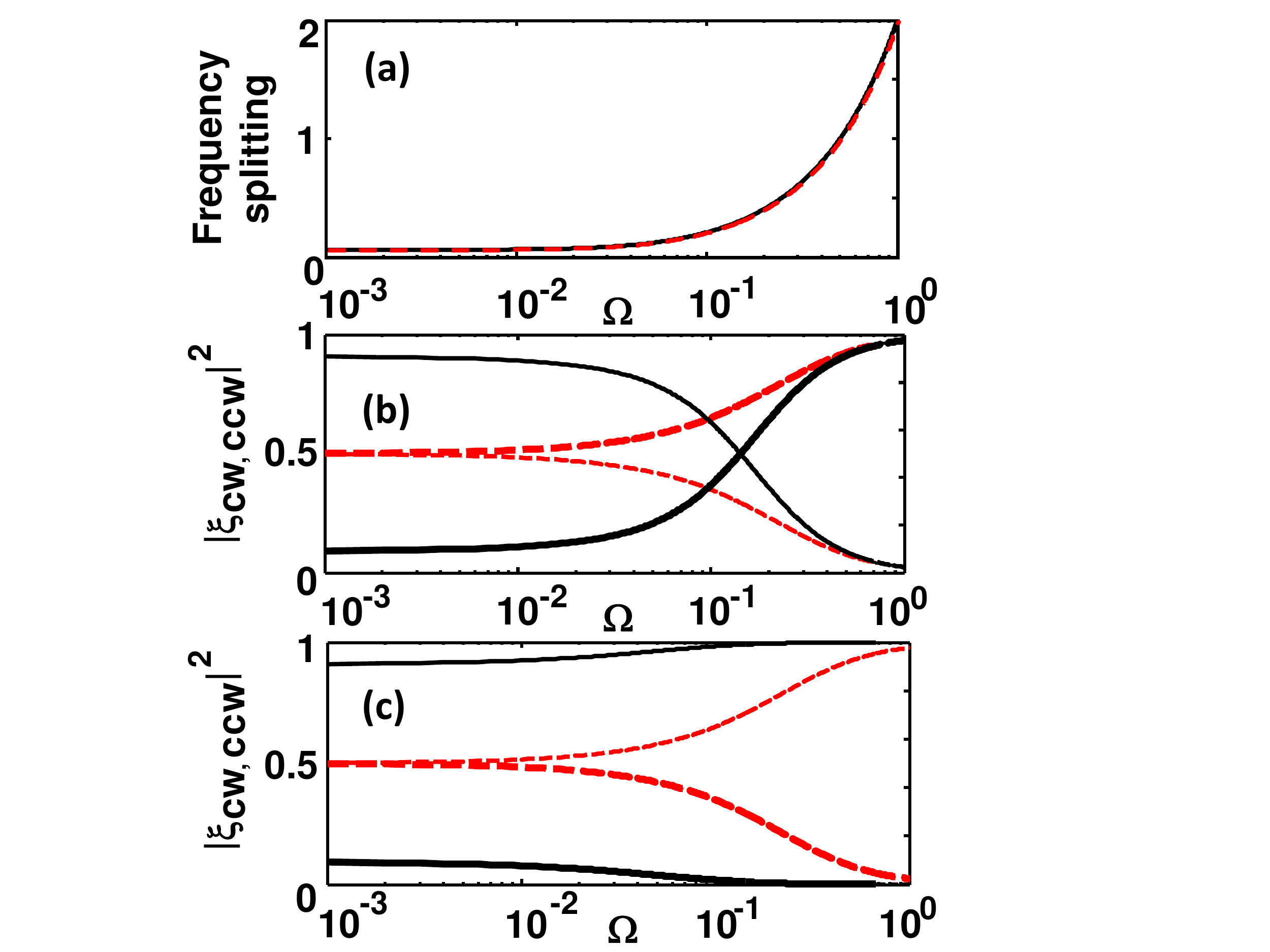}
\caption{(Color online) Comparison of Sagnac effect in a deformed microcavity with chiral symmetry ($\eta=1$, dashed line) and without chiral symmetry ($\eta=0.1$, solid line). (a) (Dimensionless) frequency splitting of the two modes as a function of rotation frequency $\Omega$. (b,c) Evolution of CW (thick line) and CCW (thin line) traveling-wave components in the quasi-degenerate modes with rotation.
}
\end{figure}

Next we investigate how the emission patterns of chiral microcavities are modified by rotation.
Without rotation, a pair of quasi-degenerate modes are expected to have similar far-field patterns, because they are both dominated by either CW or CCW traveling waves.
With rotation, one of them is changed from co-propagating to counter-propagating mode, and its far-field pattern will change dramatically if the CW and CCW waves have distinct output directions.
To illustrate this, we simulate numerically open chiral cavities.
We choose dielectric microdisks with the shape of asymmetric lima\c{c}on, which have high quality ($Q$) factor and small frequency splitting between the quasi-degenerate modes in the non-rotating frame \cite{Wiersig1}. The high $Q $ enhances the sensitivity to rotation, and the small frequency splitting reduces the dead zone.

In the polar coordinates, the boundary of an asymmetric lima\c{c}on cavity is given by $r(\theta) = R[1+\epsilon_1 \cos(\theta) + \epsilon_2 \cos(2 \theta + \delta)]$, where $R$ is the radius, $\epsilon_1$ and $\epsilon_2$ are the deformation parameters, $\delta$ sets the degree of chirality.
For $\delta = m \pi$ ($m$ is an integer), the cavity has the chiral symmetry, and the coupling from CW wave to CCW wave is equal to that from CCW to CW. As $\delta$ deviates from $m \pi$, the chiral symmetry is broken, so is the balance between CW and CCW wave coupling.
Consequently, each pair of quasi-degenerate modes are dominated by either CW or CCW wave.
Figure~2 shows a pair of transverse-magnetic (TM) modes calculated by the finite-difference frequency-domain (FDFD) method \cite{COMSOL}.
%The normalized frequencies of the two modes are around $kR \sim 6.2098$, where  $k = 2 \pi / \lambda$, and $\lambda$  is the vacuum wavelength. They are  almost degenerate with a relative frequency difference $\Delta (kR) /kR < 10^{-6}$. The $Q$ factors of the two modes are $\sim$ 56,500. They are spectrally overlapped as their linewidths exceed the frequency spacing by more than an order of magnitude.
The intracavity electric field (perpendicular to the cavity plane) is expanded in the cylindrical harmonics, $E^{(in)}_z(r, \theta) = \sum_{-\infty}^{\infty} a_m J_m(nkr) e^{i m \theta}$, where $J_m$ is the $m$-th order Bessel function of the first kind.
Positive (negative) values of angular momentum index $m$ correspond to CCW (CW) traveling wave components. %The origin of this expansion is chosen to be $(x,y) = (\epsilon_1 R/2, 0)$.
The distributions of $|a_m|^2$ in Fig. 2(a,b) illustrate that both modes consist of more CW wave components than the CCW ones.
The spatial chirality of a mode, defined as $\alpha \equiv 1 - \min \left( \sum_{-\infty}^{-1} |a_m|^2, \sum_{1}^{\infty} |a_m|^2 \right) / \max \left( \sum_{-\infty}^{-1} |a_m|^2, \sum_{1}^{\infty} |a_m|^2 \right) $, is equal to 0.25 for this pair of modes.

As shown in Fig. 2(c), the far-field patterns of these two modes are similar.
To find the output directions for CW and CCW traveling-waves, we decompose the electric field outside the cavity with outgoing harmonic waves, $E^{out}_z(r, \theta) = \sum_{-\infty}^{\infty} b_m H_m^{(1)}(kr) e^{i m \theta}$, where $H_m^{(1)}$ is the $m$-th order Hankel function of the first kind.
By summing only positive or negative $m$ terms in the field expansion and taking $r \rightarrow \infty$, we obtain the far-field intensity patterns for the CW and CCW waves separately.
As shown in Fig. 2(d), the main output direction of CW wave is $\theta \simeq 0.73 $ radians, while for the CCW wave $\theta \simeq 2.79$ radians.
Due to dominant presence of CW wave in both resonances, their far-field patterns are similar to that of the CW wave.

\begin{figure}[htbp]
\centering
\includegraphics[width=1\linewidth]
{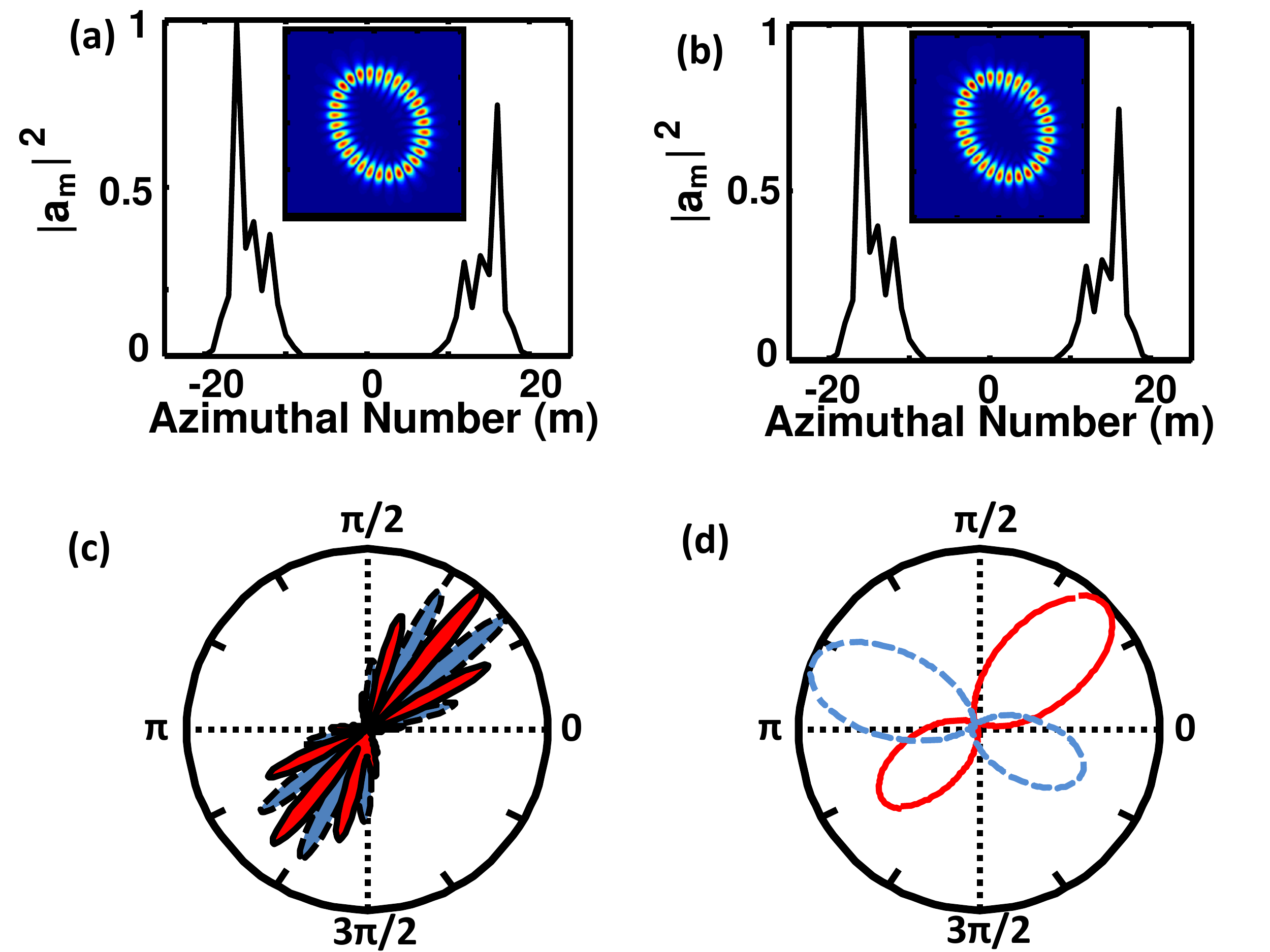}
\caption{(Color online) A pair of quasi-degenerate modes($\lambda = 598$ nm) in a non-rotating dielectric disk ($n=3.0$, $R$ = 591 nm) of asymmetric lima\c{c}on shape ($\epsilon_1 = 0.1$, $\epsilon_2 =0.075$, $\delta = 1.94$ radians).
(a,b) Field intensity distributions (inset) and angular momentum components (main panel) inside the cavity. Both modes consist of more CW wave ($m<0$) than CCW wave ($m>0$).
(c) Angular distributions of emission intensities at a distance $r = 50R$ from the cavity center. Both modes have similar output directions.
(d) Far-field patterns of CW (red solid line) and CCW (blue dashed line) waves, showing distinct output directionalities.
}
\end{figure}

Next we consider the asymmetric lima\c{c}on cavity rotating counter-clockwise with a constant angular velocity $\Omega$ around a fixed axis perpendicular to the cavity plane.
In the rotating frame where the cavity is stationary, the Maxwell equations remain the same but the constitutive relations are modified \cite{Shiyozawa1, Steinberg1, Harayama1}.
The rotation speed is slow enough that only the leading order terms of $\Omega$ are considered.
We used a finite-difference time-domain (FDTD) algorithm, adapted to the rotating frame \cite{Sarma1}, to calculate the mode profile and emission pattern.
As seen in Fig. 3, one of the two modes in Fig. 2 converts to CCW traveling-wave, while the other one remains CW, so their emission patterns become very different.
The dramatic change of the emission pattern due to rotation indicates that it might be more sensitive to rotation than the Sagnac effect in a microcavity.

\begin{figure}[htbp]
\centering
\includegraphics[width=1\linewidth]
{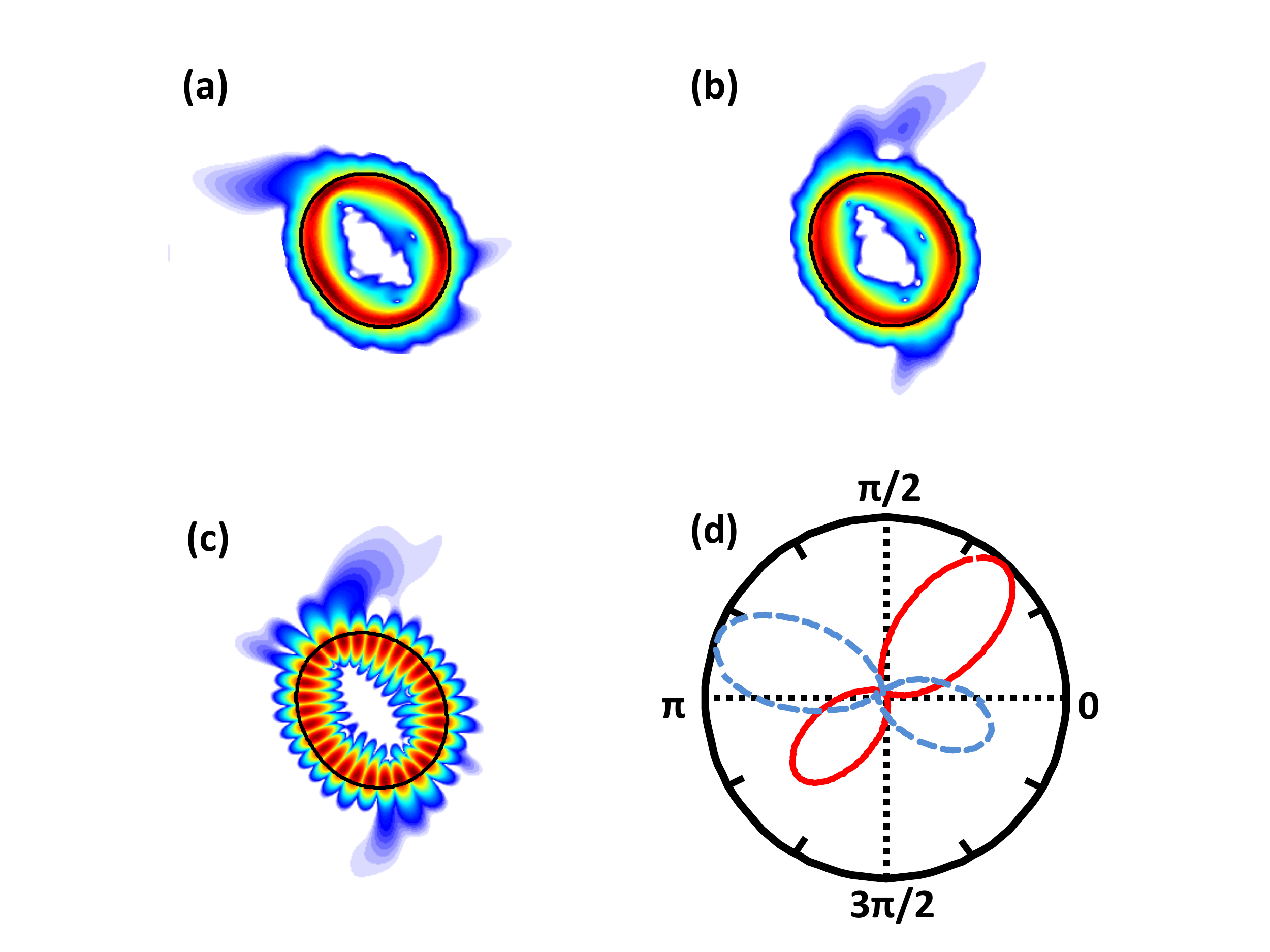}
\caption{(Color online) Emission from the asymmetric lima\c{c}on cavity in the rotating frame. (a,b) Field intensity distributions of the two modes in Fig. 2 at the normalized rotation frequency $\Omega R/c = 0.001$. The intensities outside the cavity are enhanced to illustrate the main emission directions of the two modes. (c) Field intensity distribution for one of the quasi-degenerate modes in the non-rotating frame, which evolves with rotation to the mode in (a), and the main output direction is changed dramatically. (d) Angular distribution of far field emission intensity for the two modes in (a,b).
}
\end{figure}

For a quantitative analysis, we calculate rotation-induced changes of emission intensities in certain directions. We assume the photodetectors are stationary in the rotating frame and placed at a distance of $3R$ from the cavity center. Seed pulses are launched from ten randomly chosen locations within the cavity to excite the the quasi-degenerate pair of modes in Fig. 2. After the seed pulse passes by, the photodetectors are turned on to measure the emission intensity.
Figure 4(a) plots the temporally-integrated intensity $I_e$ as a function of the emission angle $\theta$ at three rotation speeds.
The beating of emission from the two excited modes leads to oscillations of $I_e$ with $\theta$, which depend on the initial excitation condition.
As we change the rotation speed $\Omega$, we keep the initial excitation condition the same, so that we can compare the emission patterns and track their changes due to rotation.
For a quantitative comparison, $I_e(\theta)$ is normalized, $\int_0^{2 \pi} I_e(\theta) d\theta = 1$.
With increasing $\Omega$, some peaks of $I_e(\theta)$ increase and others decrease [Fig. 4(a)].
This is attributed to the evolution of the co-propagating traveling-wave resonances to counter-propagating ones by rotation [Fig. 1(b,c)].
The main emission peak at $\theta \simeq 0.7$ radians is from the CW wave, and its intensity decreases as one of the modes changes to CCW wave by rotation.
Meanwhile, the secondary peaks at $\theta \simeq 2.8$ radians increases with $\Omega$, as they are from the CCW wave.
Figure 4(b) plots the relative changes in the main peak intensity and in the ratio of main peak to the secondary peak intensities as a function of the normalized rotation speed $\Omega R/c$ ($c$ is the speed of light in vacuum).
The peak ratio is about two times more sensitive to rotation than the peak intensity.
To compare with the Sagnac effect, we calculate the frequency splitting $\Delta \omega$ of these two modes in a circular cavity of the same area as the asymmetric lima\c{c}on.
The normalized frequency splitting $\Delta \omega / \omega_0$, where $\omega_0$ is the resonant frequency in the non-rotating cavity, reflects the relative change of the resonant frequency by rotation.
As shown in Fig. 4(b), it is more than three orders of magnitude lower than the relative change in emission intensity.
Even if we increase the radius of the circular cavity to $3R$, which is equal to the distance from the photodetectors to the cavity center, the Sagnac splitting (at the same frequency) is still two orders of magnitude smaller than the change in emission pattern.
A linear fit of the data in the log-log plot of Fig. 4(b) gives the slopes, which determine the sensitivity to rotation.
The slope for the relative change in intensity of the main peak is more than two orders of magnitude larger than the slope of the normalized frequency splitting in the circular cavity with radius $3R$.

\begin{figure}[htbp]
\centering
\includegraphics[width=1\linewidth]
{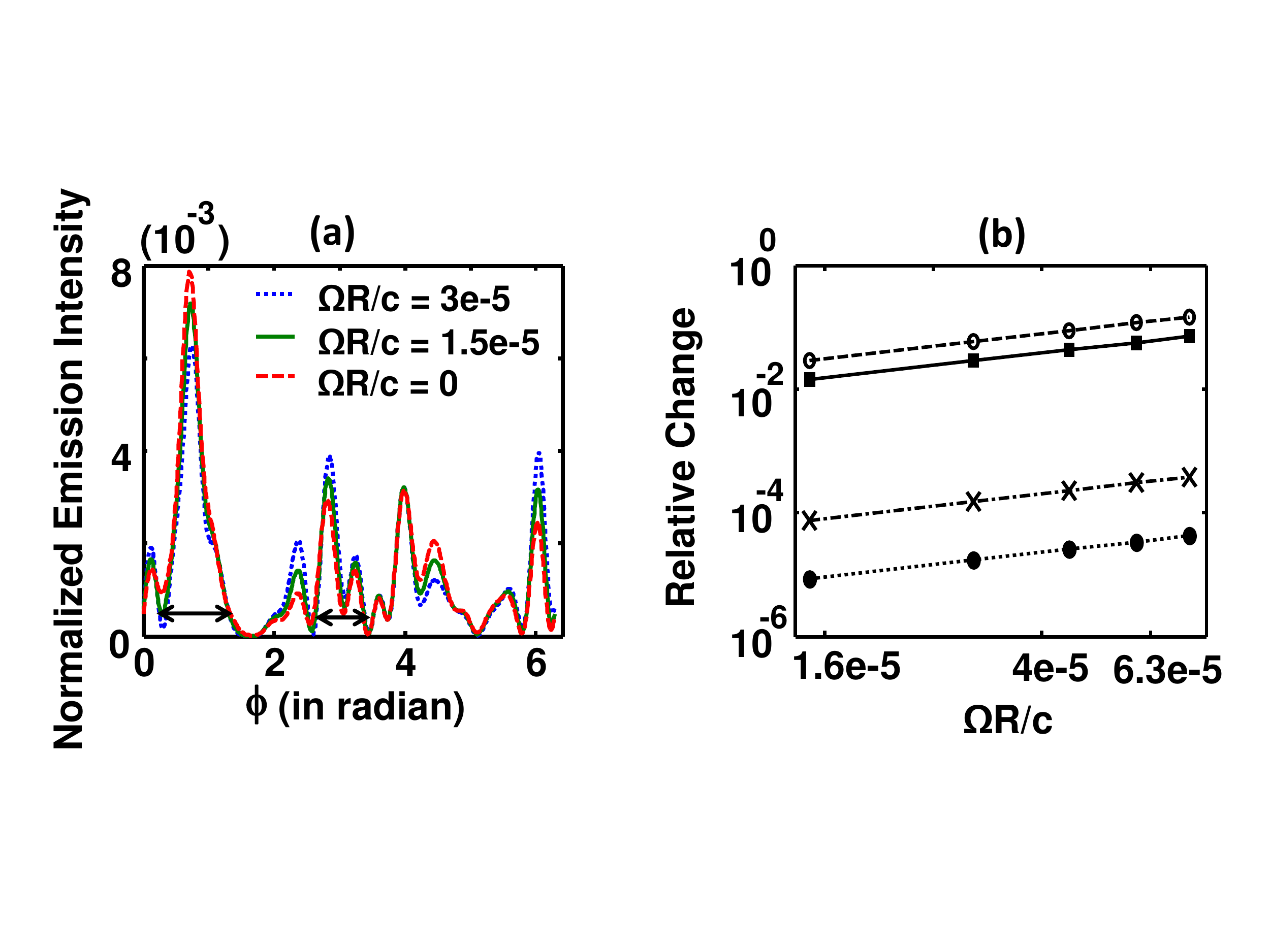}
\caption{(Color online) Rotation-induced changes of emission intensities in certain directions from the same cavity as in Fig. 3.
(a) Angular distribution of emission intensity $I_e$ at a distance of $r = 3R$ from the cavity center for three rotation speeds.
(b) Relative changes in the main emission peak intensity (at $\theta = 0.73$ radians) (solid squares, solid line) and in the ratio of main peak intensity over the secondary peak intensity (at $\theta = 2.79$ radians) (open circles, dashed line) vs. the normalized rotation frequency $\Omega R/c$. Both intensities are integrated over a range of angle marked by the double-arrowed segments in (a). For comparison, relative changes of resonant frequencies, $\Delta \omega / \omega_0$, are plotted for circular cavities with radii $R$ (solid circles, dotted line) and $3R$ (crosses, dash-dotted line).}
\end{figure}

The difference in the output directionalities of CW and CCW waves can be used to determine the direction of rotation.
Due to the breaking of chiral symmetry of the cavity shape, the quasi-degenerate modes have a preferred sense of rotation.
For example, the two modes in Fig. 2 are both dominated by CW traveling waves. One of them is changed to CCW by rotation, and its frequency is reduced (increased) if the rotation is in the CCW (CW) direction.
By measuring the frequency of emission in the direction of main output for CCW or CW wave, we can identify the direction of rotation.

To confirm the spatial chirality improves the emission sensitivity to rotation, we tune the degree of structural chirality of the asymmetric lima\c{c}on cavity and track the change of emission pattern by rotation.
As $\delta$ varies from $0$ to $\pi$, the spatial chirality $\alpha$ of the pair of modes in the non-rotating cavity shown in Fig. 2 first increases with $\delta$, reaches the maximum around $\delta = 1.94$, then decreases to zero at $\delta = \pi$ \cite{Wiersig1}.
We calculate the emission patterns for various values of $\delta$ in the rotating frame.
As shown in Fig. 5(a), the relative change of the main emission peak intensity increases monotonically with $\alpha$ at a fixed rotation speed.
To interpret this result, we compute the farfield patterns for CW and CCW waves in the non-rotating cavities with different $\delta$.
At $\delta = 0$, both CW and CCW waves emit predominantly in the direction close to $\theta = \pi/2$ [Fig. 5(b)], a slight difference of their emission directions is a result of wave effects in the wavelength-scale cavity \cite{Brandon1}.
As $\delta$ increases from $0$ to $\pi$, the main emission direction of the CW wave moves towards $\theta = 0$, while the CCW wave towards $\theta = \pi$ [Fig. 2(d)].
Meanwhile, the secondary emission peak, which is in the opposite direction of the main peak, grows monotonically [Fig. 5(c)].
The quantitative difference between CW and CCW emission patterns is characterized by $\beta = \int_0^{2 \pi} |I_{CW}(\theta) - I_{CCW}(\theta)| d \theta$, which is plotted as a function of $\alpha$ in Fig. 5(a).
Both $I_{CW}(\theta)$ and $I_{CCW}(\theta)$ are normalized, $\int_0^{2 \pi} I_{CW, CCW}(\theta) d \theta = 1$.
The monotonic increase of $\beta$ with $\alpha$ indicates the emission patterns for CW and CCW waves become more distinct at higher chirality, consequently the total emission pattern changes more significantly with rotation. The maximal spatial chirality provides the highest sensitivity of microcavity output to rotation.

\begin{figure}[htbp]
\centering
\includegraphics[width=1\linewidth]
{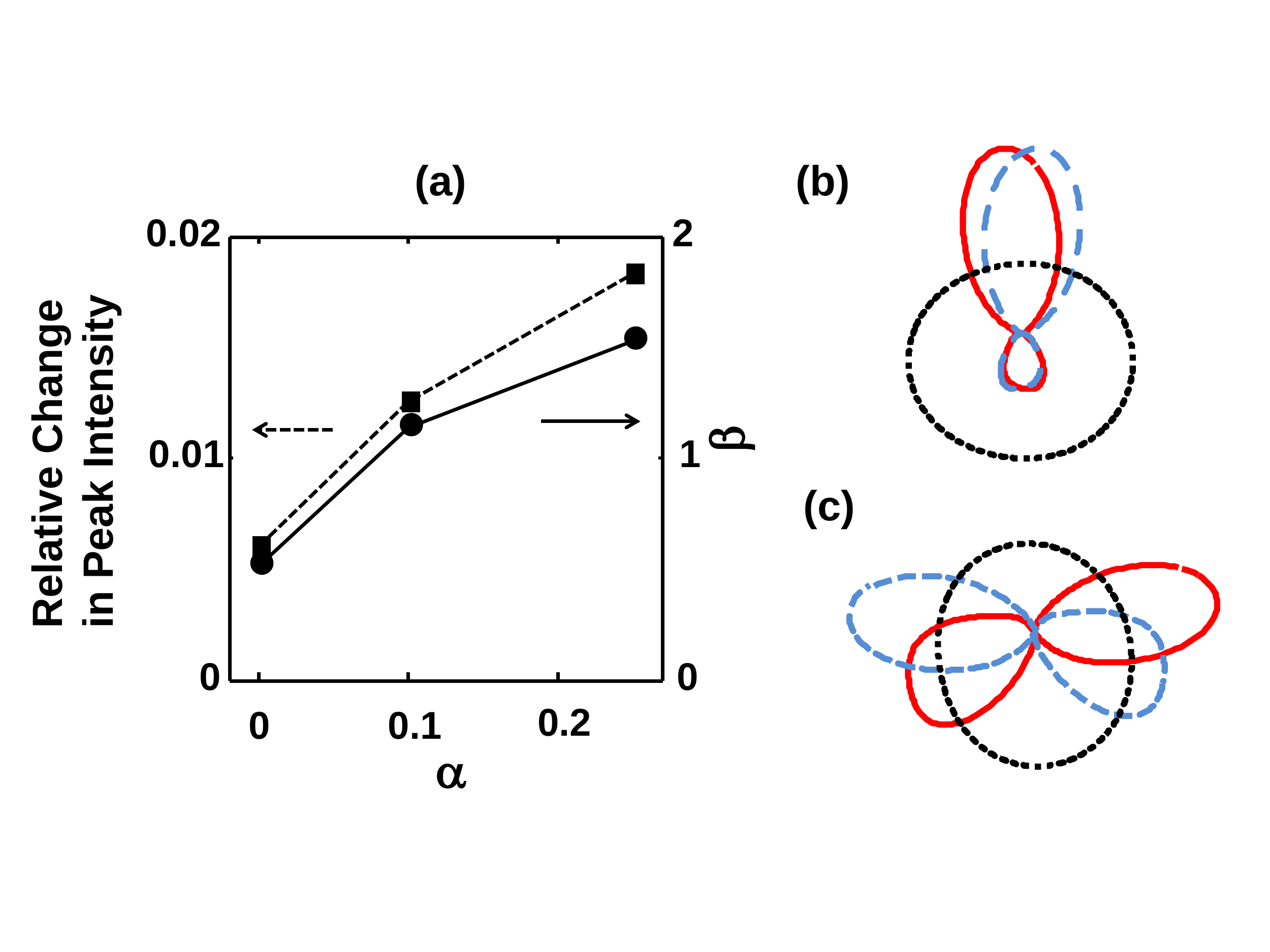}
\caption{(Color online) Tuning the spatial chirality of the quasi-degenerate modes in Fig. 2 by varying $\delta$ of the lima\c{c}on cavity. All other parameters remain the same.
(a) Relative change of the emission intensity in the main output direction (solid squares, dashed line) as a function of spatial chirality $\alpha$. The rotation frequency is fixed at $\Omega R/c \simeq 1.5\times 10^{-5}$. The difference between the emission patterns for CW and CCW waves in the non-rotating cavity is quantified by $\beta$ (solid circles and solid line), which increases with the spatial chirality $\alpha$.
(b,c) show the emission patterns for CW wave (red solid line) and CCW wave (blue dashed line) in two cavities with $\delta = 0$ (b), and $2.75$ (c). The dotted line marks the cavity boundary.
}
\end{figure}

Due to limited computing power, we simulate very small cavities, i.e. the cavity size is comparable to the vacuum wavelength. With an increase of the cavity size, the emission sensitivity to rotation will be enhanced. This is because the spatial chirality increases with cavity size \cite{Wiersig1}, along with an increase of the $Q$ factors and a decrease of the intrinsic frequency splitting (without rotation).

%In summary, we investigate how rotation modifies the resonant modes in optical microcavities with broken chiral symmetry. Without rotation a 2D chiral cavity supports quasi-degenerate pairs of co-propagating modes, which evolve to counter-propagating modes with rotation. While the frequency shift due to rotation is similar to that in a symmetric cavity, the emission pattern of a chiral microcavity changes much more significantly by rotation. The enhanced sensitivity is attributed to distinct output directions for CW and CCW waves. By tuning the degree of spatial chirality, we are able to maximize the difference between CW and CCW farfield patterns, and achieve the highest sensitivity of microcavity output to rotation. Since the rotation-induced change in emission pattern is orders of magnitude larger than the Sagnac frequency splitting, our work brings a new direction for future development of ultrasmall optical gyroscopes.

\begin{acknowledgments}
This work is funded partly by NSF under Grant Nos. ECCS-1128542 and DMR-1205307 and by DFG under Grant WI1986/6-1.
\end{acknowledgments}

 \end{document}